%% file: sn-article.tex
\documentclass[pdflatex,sn-mathphys]{sn-jnl}% Math and Physical Sciences Reference Style
\usepackage{tabularx}

\usepackage{lineno,hyperref}
\modulolinenumbers[5]
\usepackage[utf8]{inputenc}
\usepackage[T1]{fontenc}
\usepackage{graphicx}% Include figure files
\usepackage{dcolumn}% Align table columns on decimal point
\usepackage{bm}% bold math
\usepackage{epsfig}
\usepackage{subfigure}
\usepackage{graphics}
\usepackage{epstopdf}
\usepackage[english]{babel}
\usepackage{amsfonts}
\usepackage{mathrsfs}
\usepackage{amsthm}
\usepackage{amssymb}
\usepackage{makecell}

\usepackage{enumerate}
\usepackage{subfigure}
\usepackage{algorithm}
\usepackage{algorithmicx}
\usepackage{algpseudocode}
\usepackage{hyperref}

\usepackage{multirow}

\theoremstyle{thmstyleone}%
\theoremstyle{thmstyletwo}%

\theoremstyle{thmstylethree}%

\begin{document}

\title{Longitudinal Complex Dynamics of Labour Markets Reveal Increasing Polarisation}


%Shahad Althobaiti, Ahmad Alabdulkareem , Judy Shen, Iyad Rahwan, Morgan Frank, Esteban Moro and Alex Rutherford

\author[1]{\fnm{Shahad} \sur{Althobaiti}}

\author[2]{\fnm{Ahmad} \sur{Alabdulkareem}}

\author[3]{\fnm{Judy} \sur{Hanwen Shen}}

\author[4]{\fnm{Iyad} \sur{Rahwan}}

\author[5,6,7]{\fnm{Morgan} \sur{Frank}}

\author[7,8,9]{\fnm{Esteban} \sur{Moro}}

\author[4]{\fnm{Alex} \sur{Rutherford}}

\affil[1] {The Center for Complex Engineering Systems at King Abdulaziz City for Science \& Technology (KACST) and Massachusetts Institute of Technology (MIT), Riyadh, Saudi Arabia}
% https://jcep.kacst.edu.sa/center-main/ccs

\affil[2]{Intelmatix, Intelligent Informatics Technologies, Cambridge, MA, United States}

\affil[3] {Computer Science Department, Stanford University, CA, United States}

\affil[4]{\orgdiv{Center for Humans and Machines}, \orgname{Max Planck Institute for Human Development}, \orgaddress{\city{Berlin}, \country{Germany}}}

\affil[5]{Department of Informatics and Networked Systems, University of Pittsburgh, Pittsburgh, PA, United States}

\affil[6]{Digital Economy Lab, Institute for Human-Centered Artificial Intelligence, Stanford University, Stanford, CA, United States}

\affil[7]{Media Laboratory, Massachusetts Institute of Technology, Cambridge, MA, United States}

\affil[8]{Department of Mathematics \& GISC, Universidad Carlos III de Madrid, Leganes, Spain}

\affil[9]{Institute for Data, Systems, and Society, Massachusetts Institute of Technology, Cambridge, MA, USA}


\abstract{In this paper we conduct a longitudinal analysis of the structure of labour markets in the US over 7 decades of technological, economic and policy change. We make use of network science, natural language processing and machine learning to uncover structural changes in the labour market over time. We find a steady rate of both disappearance of jobs and a shift in the required work tasks, despite much technological and economic change over this time period. Machine learning is used to classify jobs as being predominantly cognitive or physical based on the textual description of the workplace tasks. We also measure increasing polarisation between these two classes of jobs, linked by the similarity of tasks, over time that could constrain workers wishing to move to different jobs.
}




\maketitle

\section*{Introduction}\label{sec1}

Economic inequality is a highly undesirable outcome for policy makers. Inaccessibility of jobs within the labour market has the potential to exacerbate this inequality by limiting opportunities for social mobility. Labour markets match skills that workers possess and tasks that firms require to be performed. How workers are allocated to jobs have many determinants including the supply and demand of worker skills and training, economic policy, geography, economic conditions, exogenous events and, increasingly, skill based technological change~\cite{National_Academies_of_Sciences_Engineering_and_Medicine2017-ry}. However, fundamentally, workers are constrained by the jobs available and the degree to which they possess the necessary skills for those available jobs.

Organisations seek to optimally and dynamically allocate workplace tasks among (i) distinct occupations such that workers possessing certain skills can be most closely matched to the tasks required of those occupations and (ii) to machines that can efficiently perform certain tasks, in order to maximise productivity. As a result, the set of available occupations tends to change over time. That is to say, some occupations may change by adding or removing tasks due to a different reallocation of those tasks across occupations or due to automation of those tasks and the replacement of human labour with machines~\cite{Tong2021-kv}. Occupations may become completely obsolete and new occupations may be created.

Consequently, it is possible that the occupations available to a worker are effectively siloed into groups with non-overlapping skill requirements. Especially if the effects of automation are likely to be complementary or deleterious heterogeneously between different kinds of jobs. In this case, a worker in a given occupation has fewer other occupations which she might reasonably be expected to perform effectively without significant retraining.

It is generally accepted that automation can lead to displacement of low skilled jobs and, increasingly, mid skilled jobs exacerbating polarisation~\cite{Brynjolfsson2011-uf}. An overarching question which remains unanswered is the extent to which current waves of automation, principally due to advances in Artificial Intelligence and to a lesser extent those in hardware, software and cloud computing, differ fundamentally from those in the past e.g. mechanical automations, personal computing and electronic communication. For example, while early skills based technological change clearly displaced physical jobs it is likely that emerging technological automations will increasingly impact white collar work. At the same time, new technology has the potential to complement the non-automated tasks of jobs leading to increased productivity. In light of this, it is extremely informative to examine past changes to the labour market, prior to recent technological change.

In light of the effects of automation which may be both complementary and deleterious~\cite{Acemoglu2018-fn}, the changing nature of labour is being increasingly scrutinised. A consistent concern since earlier waves of automation is that the beneficial and punitive effects of the substitution of human labour with machines is felt heterogeneously across society~\cite{Mokyr2015-yq}; while high skilled workers might benefit from the lower prices of goods now produced by machines, lower skilled workers may see the wages of suitable jobs drop. In addition, changes in the structure of the labour market could restrict the options of workers when looking for a future job. This is particularly relevant as increased mobility between jobs has been highlighted as a likely near term effect of automation~\cite{Bessen2020-uq}.

Given this importance, much effort has been put into both the collection and analysis of labour data to measure labour market dynamics. This has included data collected as part of national censuses as well as dedicated labour databases such as O*NET in the US and its precursor the Dictionary of Titles. In addition many private firms offer labour marketplaces which capture vast amounts of data on supply and demand of occupations and skills as well as complementary information on wages and geography~\cite{Borner2018-eg}. Despite this, the lack of availability of suitable data to investigate the effects of automation has been highlighted~\cite{Mitchell2017-ti,Frank2019-wi}.

Recent work has represented occupations and workplace tasks using network science~\cite{Dworkin2019-bq,Del_Rio-Chanona_undated-co,Mealy2018-pa}. This powerful framework has validated the explanatory and predictive power of network measures. More specifically this work~\cite{Alabdulkareem2018-as} has illuminated polarisation in terms of skills. Polarisation can be manifested in several ways in the labour market. Goos and Manning coined the term ‘job polarisation’ to describe the growth of low skilled (personal services) and high skilled jobs (professional and managerial) relative to middle skilled jobs (manufacturing and routine office)~\cite{noauthor_undated-cc}. This was likewise observed by Autor~\cite{Autor2006-ho}, and in both cases this change was attributed to skills based technological change~\cite{Autor2013-md,Goos2014-ea}. This job polarisation is typically seen through the lens of employment share or wage share. However, network science has illuminated polarisation in other dimensions; median wage, education level and susceptibility to automation in both tasks and occupations~\cite{jobscape}.

Network science represents distinct occupations as nodes, linked by the degree of similarity between those occupations in terms of their required tasks and other attributes. Polarisation between distinct occupations can be naturally measured as the degree to which the nodes in the network are equally connected. Such polarisation is undesirable as it constrains the ability of a worker in one occupation to transition to another arbitrary occupation~\cite{Del_Rio-Chanona_undated-co}. Note that polarisation of the network of occupations is concerned with structural constraints and not the demand for specific jobs nor their wage. Nevertheless, the structural polarisation lowers resilience to changes in economic conditions due to skill based technological change, demographic change or migration. Recent work has analysed in detail the consequences of a lack of accessibility between occupations in the face of imminent disruption from automation~\cite{noauthor_undated-an} and a shift to green technologies~\cite{climateaction}.

Empirical work examining the changing nature of work can be challenging due to the long time scales for technological change to become integrated into occupations and organisations. But notable work has been done in this direction, uncovering long term trends in wages~\cite{Kelly2018-rc}, high skilled jobs~\cite{noauthor_undated-cc} and employment through patents~\cite{webb}.

The Dictionary of Titles has, in this context, been used as a limited source of structured data on job tasks. Each edition includes a ranking of each occupation on categorical scales that are not consistent across all editions. The scales measure the degree to which an occupation interacts with ‘People’ and ‘Machines’ in different ways. Previous work in this direction includes measurement of the influence of geography on the prevalence of workers in new occupations~\cite{Lin2011-il} and changing task requirements attributable to computerisation~\cite{Autor2003-vz}.

Despite this, progress in understanding the dynamical nature of work has been limited primarily by the availability of data, or more specifically the dearth of \textit{structured} data. Additionally, suitable tools and methods to analyse this data have also been lacking.

In this work we address both these aspects. In particular, our contributions are (i) the public release of a large dataset of unstructured and structured data describing occupations over 7 decades (available at ~\cite{DVN/DQW8IP_2022}) (ii) the quantification of how occupations have changed and become obsolete in this time period (iii) the application of natural language processing and machine learning to accurately categorise occupations based on unstructured data in the form of free text job descriptions and (iv) to measure and validate a trend of increasing polarisation between cognitive and physical occupations that predate recent advances in skill based technological change.

This work is further distinguished from previous work measuring the changing share of wages in high and low skilled jobs, as we are concerned with the underlying structure of available occupations and their accessibility between one another which constrains worker mobility. Our overarching motivation is to better understand how the relative accessibility of occupations has changed in the past, in order to contextualise observed trends in labour dynamics that may be attributable to skills based technological change in the present day.

\section*{Results}
In this work we make use of the Dictionary of Titles, the US Bureau of Labour Statistics precursor to the Occupational Network (O*NET) database. The Dictionary of Titles was a comprehensive catalogue of all occupations found in US workplaces with semi-regular updates every 10-16 years between 1939 and 1991. We extract the free text descriptions of the tasks of the worker as well as related meta-data through manual transcription. Each edition contains a similar order of magnitude of occupations and other summary statistics of each edition is shown in Table 1.

\begin{figure}[htbp]
	\centering
	\includegraphics[scale=0.42]{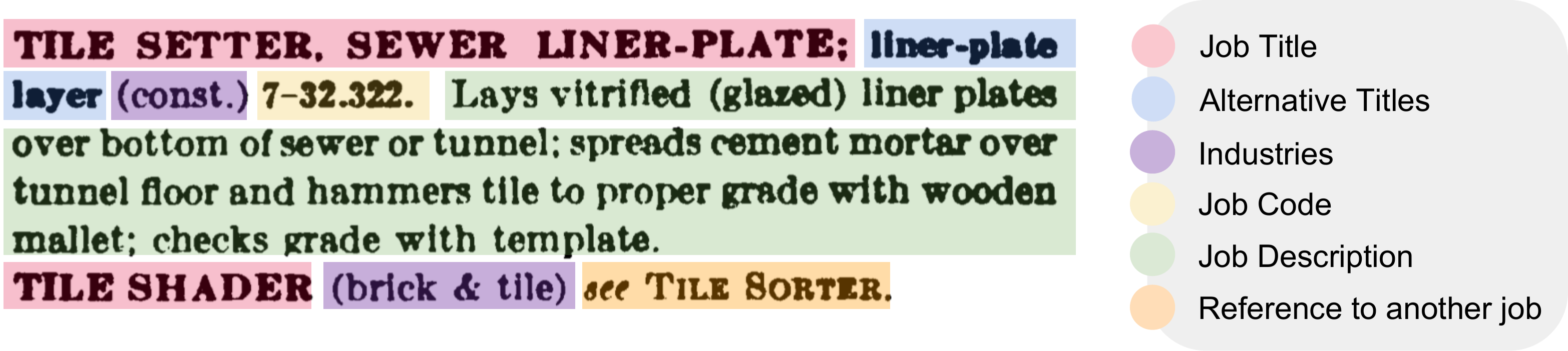}
	\caption{Example excerpt from the 1939 Dictionary of Titles, extracted data and meta-data is highlighted
}\label{dot_example}
\end{figure}

\begin{table}[]
\begin{tabular}{|c|c|c|c|c|c|}
\hline
\multicolumn{1}{|l|}{\textbf{Year}} &
  \multicolumn{1}{l|}{\textbf{\thead{Total Number \\of Occupations}}} &
  \multicolumn{1}{l|}{\textbf{\thead{Number of \\Occupations with \\Distinct Descriptions}}} &
  \multicolumn{1}{l|}{\textbf{\thead{Mean  \\Description \\Length}}} &
  \multicolumn{1}{l|}{\textbf{\thead{Mean \\Description \\Length \\(No Stop word)}}} &
  \multicolumn{1}{l|}{\textbf{\thead{Vocab \\Size}}} \\ \hline
1939 & 27,539 & 15,849 & 50.92 & 40.96 & 77,154 \\ \hline
1949 & 38,318 & 17,470 & 58.72 & 47.63 & 97,694 \\ \hline
1965 & 29,856 & 11,312 & 91.12 & 76.16 & 87,970 \\ \hline
1977 & 12,625 & 11,996 & 95.78 & 79.81 & 79,601 \\ \hline
1991 & 13,315 & 12,728 & 99.18 & 82.7  & 83,768 \\ \hline
\end{tabular}
\label{dot_table}
\caption{Table of summary statistics of each Dictionary of Titles Edition}
\end{table}

\begin{figure}[htbp]
	\centering
	\includegraphics[scale=0.85]{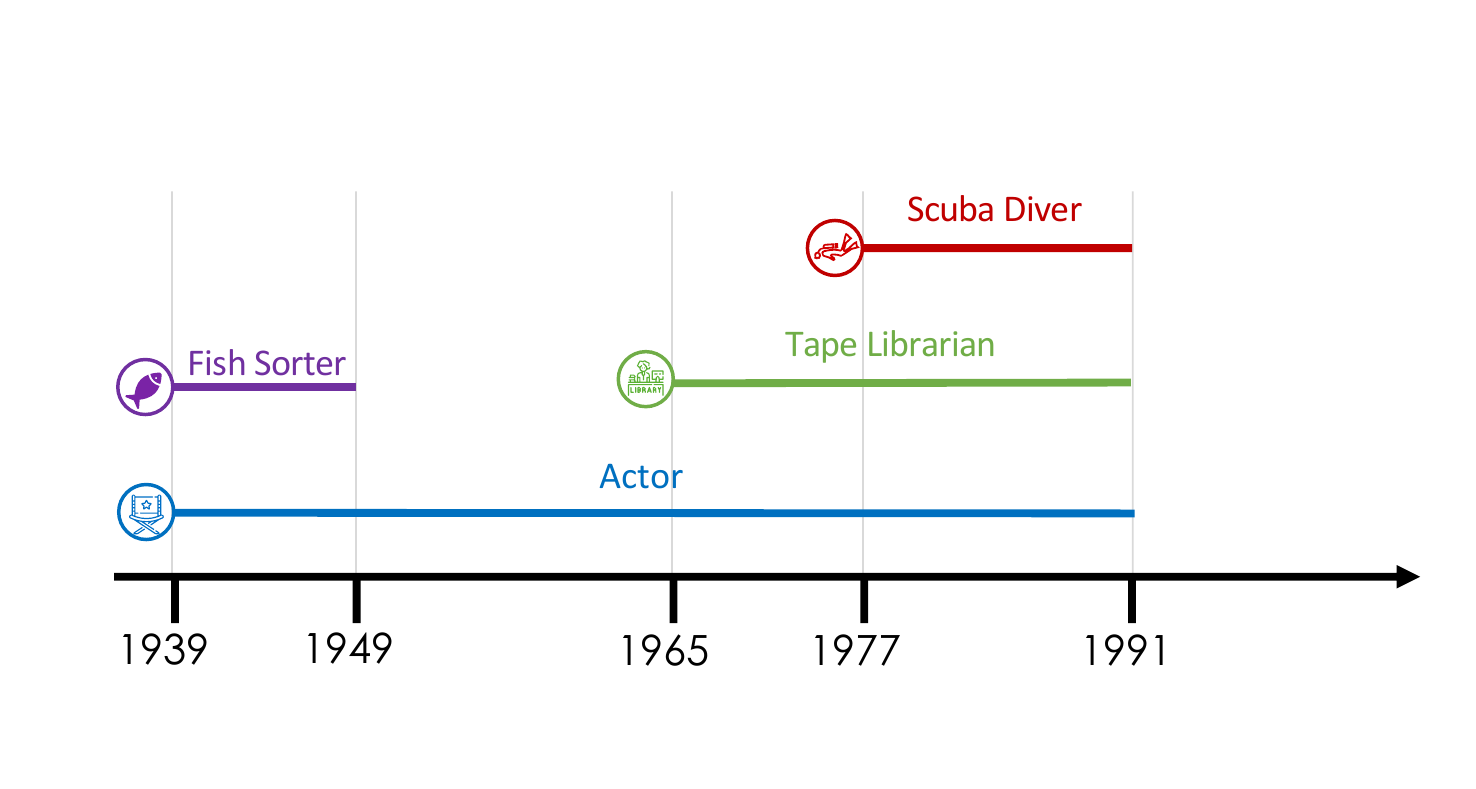}
	\caption{The presence of indicative jobs between DoT editions
}\label{in_out}
\end{figure}

We begin by measuring the degree to which occupations become obsolete i.e. \textit{an occupation is found in one edition and not found in a subsequent edition}. For example the job titles of `Tutor`, `Actor` and `Underwriter` were all present in each edition, whereas the titles of `Media Clerk` and `Tape Librarian` were not present in the first two editions (1939 \& 1949) but were then present in the final three editions (1965, 1977 \& 1991) see Fig \ref{in_out}.

\begin{figure}[htbp]
	\centering
	\includegraphics[scale=0.10]{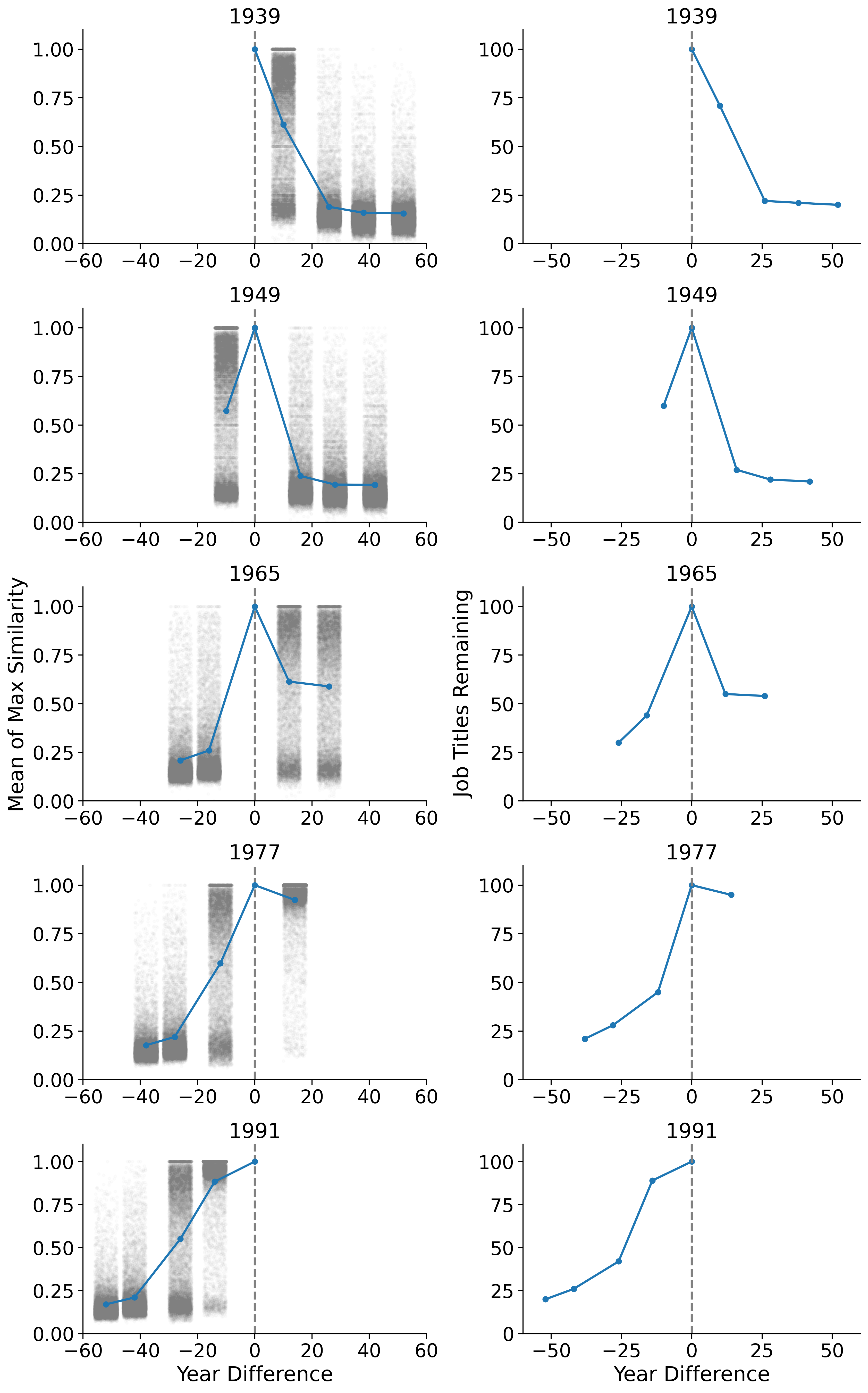}
	\caption{The maximum textual similarity between a focal occupation and all other occupations (left). Each individual occupation-wise maximum similarity across all other occupations is marked as a grey point jittered in the x-direction, the mean of the maximum similarity is marked in blue. The percentage of occupations persisting (right) in each DoT edition compared to all other editions as a function of the time difference between editions.
}\label{changes}
\end{figure}

Figure \ref{changes}, right hand column, shows the percentage of occupational titles in a focal year that are not found in other years, as a function of the time difference (which lies in the range [10-16] years). Strikingly, only a small percentage of occupations present in one edition are present more than one edition earlier or later. In fact the percentage drop can be explained quite accurately with a linear regression (r = 0.852, $p<10^{-3}$, see SI) demonstrating that 1.67\% of job titles disappear or appear per year on average.

A distinct mechanism is whereby an occupation is simply \textit{renamed}, or \textit{the majority of the tasks of one occupation now appear in a different occupation with a different name}. This can be proxied by observing the maximum textual similarity between the free text descriptions of a focal job (the green text in our example shown in Fig \ref{dot_example}) and all other jobs in a given year. In order to capture deeper semantic similarities between job tasks, that may be missed by exact word matches, we make use of a word embedding to calculate the occupation-wise similarities we report on in the main text.

Again a linear model based on the time difference has considerable explanatory power (r = 0.842, $p<10^{-3}$, see SI) predicting that the mean of maximum occupational similarities drops by 0.017 per year.

In order to construct a network between occupations for each edition of the DoT, we consider again the similarity between the descriptions of the tasks of each occupation. Our results are largely robust to various different measures of text similarity (see SI). Thus we derive 5 networks, one for each edition, from the DoT data in which nodes are occupations which are connected by links with a weight that is determined by the similarity of the text used to describe both occupations.

The significance of the structure of the network is that it describes the constraints that a
worker has in moving from one job to another job. That is, a worker is unlikely to be able to transition from one job to another arbitrarily chosen job as there will be little overlap in the skills required to perform the workplace tasks. The extent to which a network is clustered and polarised determines the degree to which all jobs \textit{are} or \textit{are not} mutually accessible for all workers. The extent to which occupations are clustered together, determines how hard it will be for a worker to transition from a similar job to a different job i.e. lower waged and skilled to higher waged and skilled. Note that the network of distinct occupations does not contain information about wages or demand for jobs in those occupations.

We seek to measure the degree of polarisation between cognitive and physical jobs in each DoT network. Community detection is a commonly used tool to exploit the polarisation or modularity in a network. Community detection decomposes a network into clusters of nodes sharing a common class or category, such that nodes are highly connected within clusters but less connected between clusters~\cite{Fortunato2016-ll}. Community detection is an optimisation process which \textit{assigns} labels from an arbitrary number of classes in order to best partition the nodes into clusters by maximizing polarisation.

Consequently, rather than applying community detection to the DoT job networks in order to uncover likely occupational clusters, we seek to classify each node in the network as requiring either predominantly \textit{physical} or \textit{cognitive} skills. Once the nodes of the network have been assigned a label i.e. the occupations in our network have been categorised as cognitive or physical, we can measure the emergent polarisation of the occupations by the degree to which the labeled nodes connect preferentially to others of the same class i.e. physical to physical or cognitive to cognitive. A large modularity value implies that cognitive jobs are only connected to other cognitive jobs and vice versa for physical jobs. Whereas a smaller value implies that cognitive and physical jobs are equally likely to be connected to one another. Our hypothesis is that the measured modularity of these networks increases over time, indicating that a physical job is less likely to be connected to a cognitive job and thus that a worker in a physical job is less able to transition to a cognitive job (and vice versa).

\begin{figure}[htbp]
	\centering
	\includegraphics[scale=0.50]{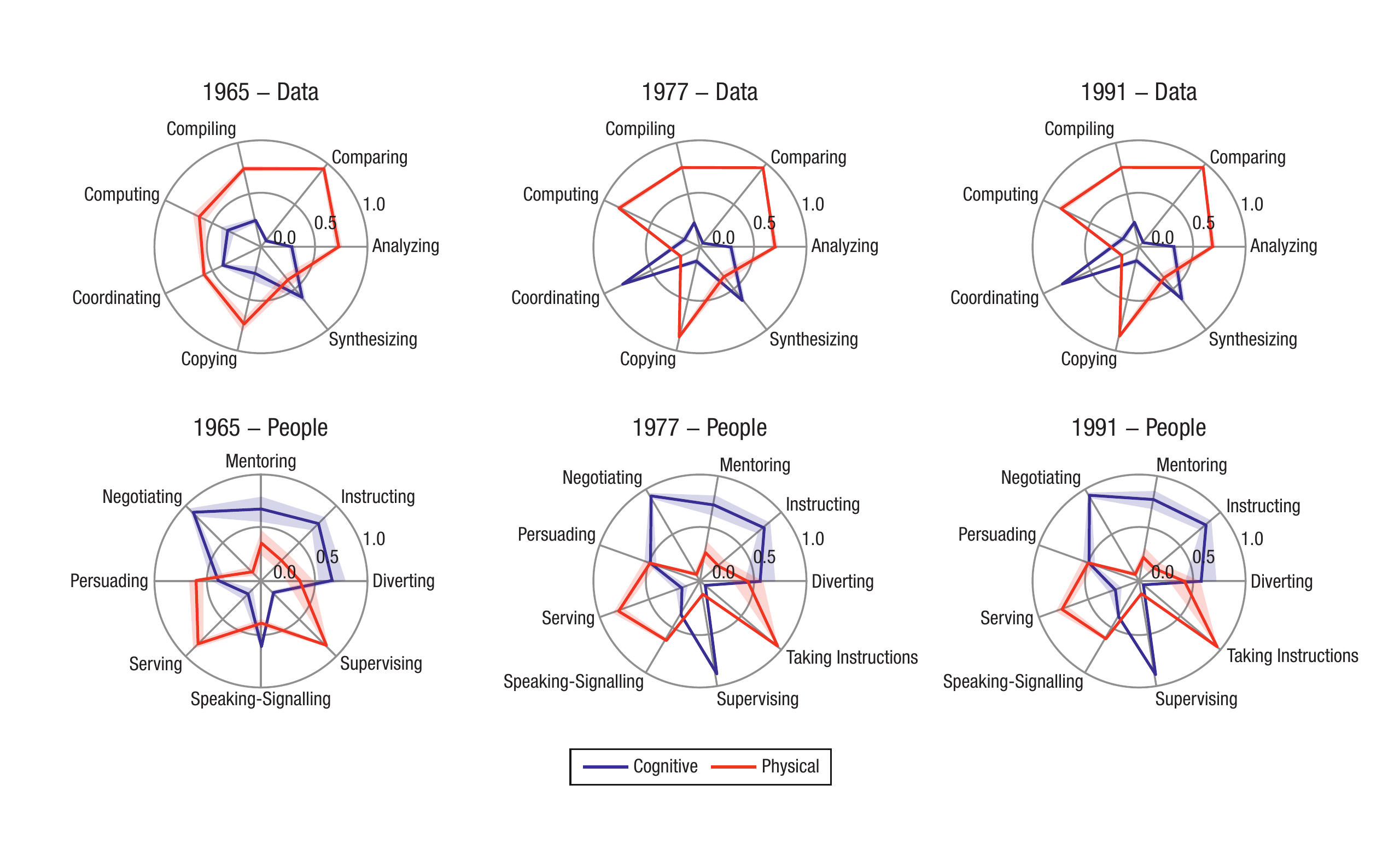}
	\caption{Glyphs showing the scores of jobs classified as physical (red) and cognitive (blue) on various measures of Data, People and Things for (a) 1965 (b) 1977 and (c) 1991
}\label{glyphs}
\end{figure}

We proceed to label the jobs in our networks by building a classification model. We train our classifier using training data from the 2018 O*NET and classifying jobs as physical or cognitive following the method in~\cite{jobscape} (see SI for details). After manual validation, our best performing classifier achieves an accuracy of 87\% (see SI for details and comparison of various classification techniques). We also validate our classification against structured data on work tasks provided in the DoT (see Fig \ref{glyphs}). The DoT editions for 1965, 1977 and 1991 all score each occupation on various interactions with People, Data and Things on an ordinal scale. For example \textit{`Instructing People`}, \textit{`Analysing Data`} and \textit{`Operating Things`} respectively. Fig \ref{glyphs} demonstrates that the occupations that are classified as physical have strikingly different task profiles in comparison to cognitive jobs. This trend is not so clear in the case of \textit{Data}, partly attributable to numerous missing values. The equivalent figures are available in the SI. We also find that physical jobs begin as the strongly majority class, reducing over time to become more balanced with cognitive jobs (see SI) consistent with previous empirical studies~\cite{Autor2003-vz}.

\begin{figure}[htbp]
	\centering
	\includegraphics[scale=0.50]{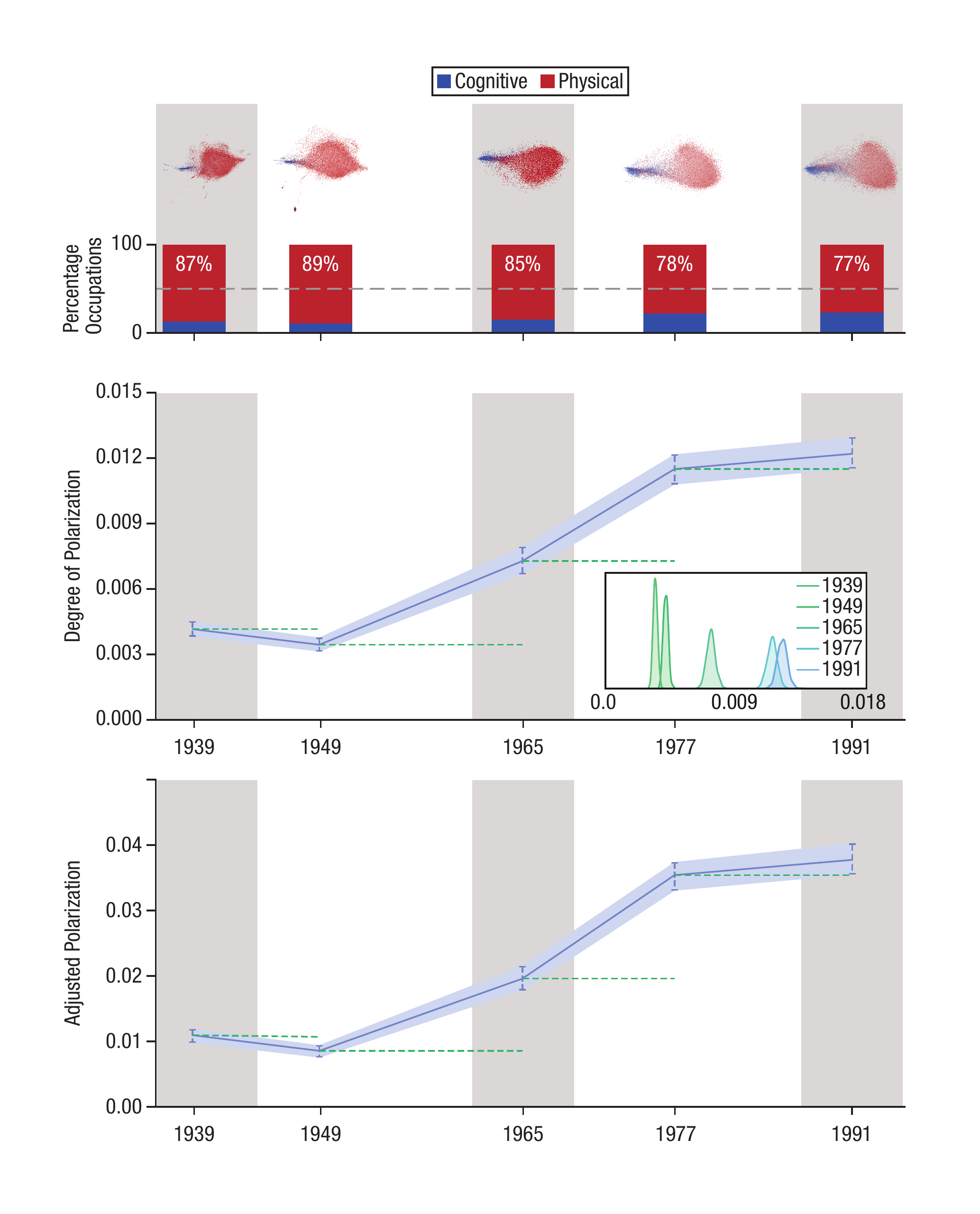}
	\caption{The degree of polarisation in each edition of the DoT. (Top) The structure of the job networks for each year. Each node is a job title and the strength of the links between job nodes is determined by the similarity of the text in the job descriptions.The layout of the networks is calculated using the Force Atlas algorithm. (Middle) The proportion of cognitive and physical jobs in each edition. (Bottom) The polarisation in the job network of each edition. The blue shading indicates confidence intervals from bootstrapped perturbations of the original network. The inset shows the distributions of these bootstrapped polarisation values for each year as kernel density estimates.}\label{polarisation}
\end{figure}

We find clear evidence of increasing polarisation~\cite{Blondel2008-bz} in DoT editions over time (Fig \ref{polarisation}B). Our results are robust to various measures of text similarity and classification techniques (see SI). The error bars represent 1,000 bootstrapped polarisation calculations based on perturbed networks in which around 35\% of nodes are removed and the total network size maintained by duplicating others. That polarisation is measured to be increasing is more remarkable, as it is despite a trend towards equal proportions of physical and cognitive jobs which would lead to \textit{lower} polarisation with all else equal. We adjust the polarisation observed in light of these changing proportions; effectively controlling for the imbalance between physical and cognitive jobs. We compare the observed, empirical polarisation in each edition to a regular network of equivalent imbalance of node classes using numerical approximations (see SI). The adjusted polarisation is also presented in Fig \ref{polarisation}C.

\section*{Discussion}

In this work we analyse unstructured data with a longitudinal dataset contained within the Dictionary of Titles to investigate changing labour market dynamics. Specifically our contributions are fivefold; (i) we present a longitudinal dataset of transcribed occupational descriptions and metadata (ii) we quantify the rate at which jobs become obsolete and change their nature over time (iii) we calculate semantic similarity between descriptions to construct networks of jobs linked by task similarity (iv) we classify jobs as physical or cognitive and validate this classification (v) we measure the polarisation of cognitive and physical jobs over time to find this has increased.

It is necessary to highlight two aspects of job market polarisation that are out of scope for this paper. Firstly, we do not examine changing employment share i.e. we do not incorporate the number of workers in each occupation. Secondly, we do not consider wages of jobs and polarisation of wages. This data is imperfectly reflected in the census from the overlapping period and to some extent can be incorporated, which we leave for future work.

The increasing polarisation observed in the job network over time is consistent with a persistent role played by new technology in reallocating workplace tasks, particularly routine tasks~\cite{Autor2008-pl}, from workers to machines. However our analysis is not able to fully disentangle exactly the causes for these structural changes, as changing workplace practises, labour regulation, offshoring and organisational restructuring might also be responsible for the changes observed~\cite{Goos2014-ea}. However we have evidence that, regardless of the causal driver, polarisation between physical and cognitive jobs has in fact been increasing steadily over half a century. While an increasing share of that trend might be attributable to technological innovation, the increase in the degree of job polarisation has been steady rather than disruptive.

\section*{Acknowledgements}
We thank Sarah Oberstetter for support on formatting figures.


\bibliography{paperpile}% common bib file

\include{main_content}
\end{document}

%% file: main_content.tex
\maketitle
\pagebreak
\setcounter{figure}{0}

\section{Data Overview}
This section will provide a description and overview of the datasets used in this paper, namely the Dictionary of Occupational Titles (DoT) and O*NET.
\subsection{Dictionary of Occupational Titles (DoT)}

In 1939, the United States Department of Labor published its first edition of a comprehensive catalogue of jobs in the US labor market. The collection of job definitions included observations from job analysts who visited workplaces and recorded the tasks for a specific job as unstructured text. For example:

\begin{quote}`\textit{BENCH-LATHE OPERATOR (mach. shop). A general term applied to a worker who operates a small lathe mounted on a workbench. Usually specifically designated according to type of lathe, as Engine-Lathe Operator; Speed-Lathe Opbbator.}'
\end{quote}
DoT also relied on information from various sources such as libraries, employers, trade and labor associations, labor organizations, and public employment offices. United States Department of Labor published five different editions of DoT (1939,\,1949,\,1965,\,1977 and 1991). In 2010 the DoT was replaced by O*NET. DoT encapsulates information about the skills and abilities presented in a specific job through its description. It also provides an occupational classification where each job is assigned to or refers to an occupational code. Each job is also associated with industrial designation, which refers to the industry or industries in which the job is based.

\begin{figure}
    \centering
    \includegraphics[scale=0.4]{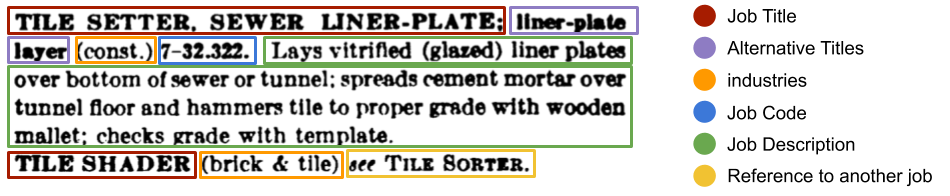}
    \caption{Example of an entry for an occupation in the 1939 DoT with data and meta-data labeled}
    \label{fig:dot2}
\end{figure}
\subsubsection{Data Collection and Transcription }
The original publications exist only in scalar images with only small sections of certain years available in a structured format. Using human transcribers we have converted this data into digital text. To validate the accuracy of the transcribers, we built a spell checker validating each word in each DoT dictionary edition. Table \ref{tran_accuracy} summaries the obtained accuracy for each DoT edition. All of the editions scored above 99.5\% accuracy, which is sufficient for our analysis.

\begin{table}[ht]
  \begin{tabularx}{\linewidth}{XXX}
    \toprule DoT Edition & No. of misspelled words         & Accuracy rate\\
    \midrule
    1939                      & 5,497        & 99.53\%\\
    1949                     & 7,536        & 99.54\%\\
    1965                         & 8,118       &99.49\%\\
    1977                       & 5,667        &99.63\%\\
    1991                       & 5,652       &99.69\%\\
  \bottomrule
  \end{tabularx}
  \caption{Summary of the transcribers' accuracy rate for each DoT edition}
  \label{tran_accuracy}
\end{table}

\subsubsection{Parsed Data and Statistics}
The entry for each job was transcribed as free text and subsequently parsed to recognise structured meta-data. Figure \ref{fig:dot2} shows the patterns that were captured by the parser including title, job code, description, industry and when a job description is the same as another job it is referenced to that job using some keywords as \emph{see or ref. to job x}. As example for the regular expression to capture titles, it consider a capitalized sentence or word that is followed by a bracket. The data was structured into job title, alternative titles, job code, description, and industries. Table \ref{parsed} provides a summary of the statistics of the parsed data. In our analysis, we only used jobs with unique descriptions and eliminated the jobs with references that don't have a description of its own and jobs with duplicated descriptions.
\begin{table}[ht]
  \begin{tabularx}{\linewidth}{XXXXXXX}
    \toprule DoT Edition & Vocab size
& No. of jobs &No. of jobs after elimination & No. of jobs with references & No. of coded jobs & Occupational Groups\\
    \midrule
    1939&   77,154    &27,539 &15849  &11,688 &5,635  &N/A\\
    1949&   97,694    &38,318 &17470  &20,843 &13,708 &N/A\\
    1965&   87,970    &29,856 &11312  &18,529 &10,885 &N/A\\
    1977&   79,601    &12,625 &11996  &N/A    &12,625 &629 \\
    1991&   83,768    &13,315 &12728  &N/A    &13,319 &640 \\
  \bottomrule
  \end{tabularx}
  \caption{Summary of the parsed data}
  \label{parsed}
\end{table}

\subsection{Occupational Information Network (O*NET)}
The Occupational Information Network (O*NET) is a publicly available resource that replaced the DoT. It started its first release in 1998, and since then, it provided an annual update on the skills and requirements of occupations. O*NET contains information detailing worker-related information such as interests,  education, skills.  In addition, it includes job-related information such as job titles, descriptions, tasks, and tools and technologies used. O*NET includes around 1100 consolidated occupations  as opposed to the DoT which provided a separate entry for every job title numbering in the thousands\footnote{\url{https://www.onetonline.org/}}.

\section{Methodology}
\begin{figure}
    \centering
    \includegraphics[scale=0.5]{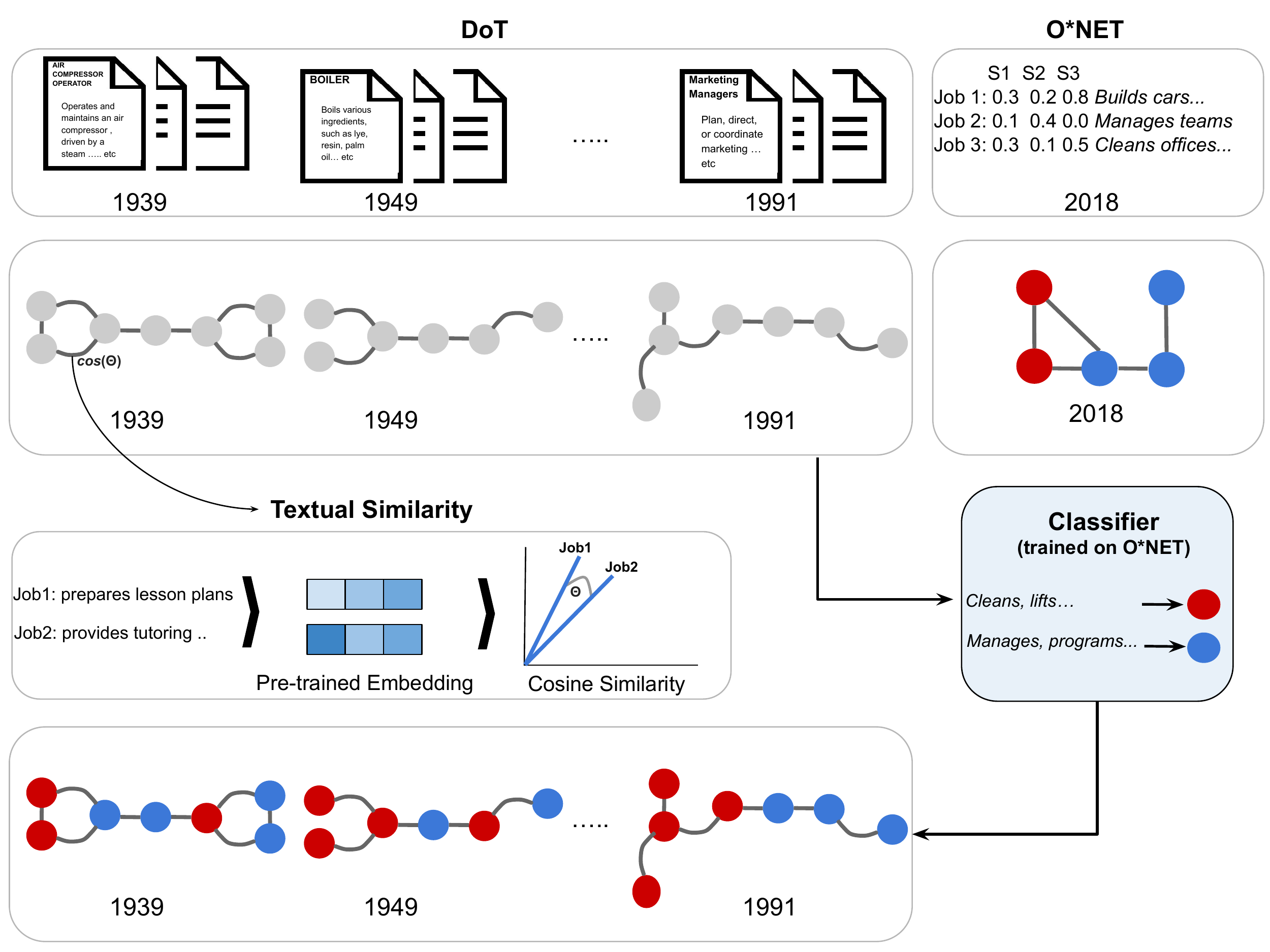}
    \caption{Demonstration of the Methodology}
    \label{fig:DoT_method}
\end{figure}

In order to extract a measure of polarisation from each edition of the DoT, we use techniques from network science, machine learning, and natural language processing. Figure \ref{fig:DoT_method} outlines our procedure. In order to quantify the value of polarization, we looked into the strength of the relationship between occupations within each DoT edition. The strength is measured using a textual similarity between job descriptions that describes the skills and tasks of the job resulting in an adjacency matrix for each DoT edition. Furthermore, we used the Louvain modularity to calculate the degree of polarization. Since we are looking for polarization, we are generally looking for two partitions of the data, but the Louvain modularity community detection tries to find the best number of partitions that maximize the modularity value.

\begin{figure}
    \centering
    \includegraphics[scale=0.7]{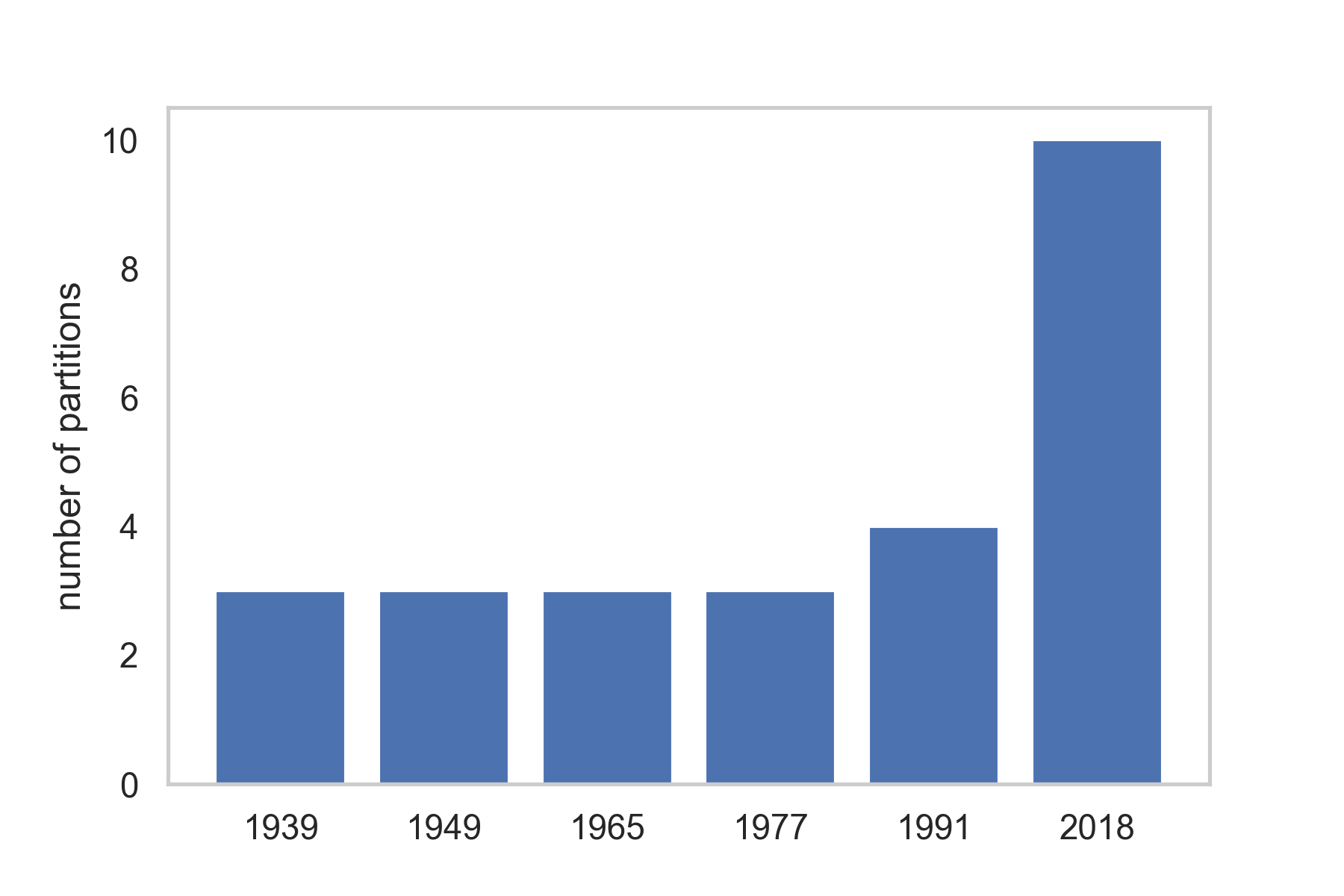}
    \caption{Bar Chart of modules discovered by Louvain modularity}
    \label{fig:louvain_clusters}
\end{figure}

Therefore, we built a binary classifier that takes a job description as an input and outputs a label indicating whether the job is dependant on sensory-physical or socio-cognitive skills and deployed our classifier for all the DoT editions. In this section, we describe the methods used in detail.

\begin{table}[ht]
  \begin{tabularx}{\linewidth}{XX}
    \toprule Model & Accuracy\\
    \midrule
    ELMO                      & 87\%      \\
    BERT                     & 85\%        \\
    RoBERTa                         & 84\%       \\
    TransformerXL Embeddings & 84\%   \\
    Stacked Embeddings(Bert, RoBERTa and ELMo) & 91\%   \\
    Ensemble model & 87\%   \\
  \bottomrule
  \end{tabularx}
  \caption{Summary of the results of the State of the Art Models}
  \label{mod_accuracy}
\end{table}

\subsection{ Textual Similarity }
The textual similarity is a technique used to calculate the similarity percentage between two strings. Similarity can be based on terms (tokens) or the semantic meaning of a sentence. In our work, we used pre-trained word embedding and cosine similarity to find the semantic similarity between jobs within each DoT edition. In our paper, we preferred to apply the semantic-based textual similarity as two jobs might have different tokens, but they are fundamentally the same. For instance, jobs like teachers and instructors in the token-based similarity would have a similarity zero, while the semantic similarity would have a closer similarity score. We used Facebook's pre-trained word embedding model called fasttext. For each job description, we find the semantic vector by averaging the word vectors of its collection of words. And a similarity score is calculated by finding a cosine similarity of two job embedding vectors. For each job in a DoT edition, we calculate the similarity with every job in the same edition based on the job description. As a result, we have an adjacency matrix with the similarity between jobs for each DoT edition. The process of obtaining the semantic similarity is visually demonstrated in figure \ref{fig:part1} part B.

\subsection{ Polarization Calculations} \label{polarization}
For calculating the polarization, we used the Louvain modularity optimization function. Given a representation of weighted network (adjacency matrix) and the binary partition, Louvain modularity function measures/calculates the strength of connections inside a community (intra-connections) compared to connections between communities(interconnections). To capture the polarization trend, we want to test if the intra-connections is becoming stronger over time as opposed to the interconnections overtime.

\subsubsection {Bootstrapping}

We then used the bootstrapping technique on the jobs in the adjacency matrix to estimate the degree of polarization. Resampling is done on the jobs with replacement by deleting and duplicating a job entry. After each sample is generated, we calculate the polarization value; the process was repeated 1000 times. As a result, we have a sample of 1000 polarization values for each DoT edition.

\subsection {Classification}
In order to calculate the polarization, we need two main categories that represent the nature of polarization , namely sensory-physical and socio-cognitive. To achieve that, we built a binary text classifier that takes a job description and labels it as sensory-physical or socio-cognitive. The model is trained on balanced data of 672 O*NET job descriptions that were obtained from job space mentioned in the related work. The data is split into 80\% for training and 20\% for testing and validation. We used both traditional and advanced methods for training and testing to provide the most robust result for our analysis methods will be described in detail in the following section. After obtaining the best classifier, all the jobs in the DoT edition were labeled to describe their nature of tasks (sensory-physical or socio-cognitive). For validating the feasibility of applying the classifer to DoT descriptions, we tested the classifier's results against the human-based labeling of a sample of DoT jobs.

\subsubsection {Traditional Methods}

Traditional methods include using traditional machine learning algorithms and features to build the text classifiers. We built a baseline model based on features and algorithms that are widely used in text classification tasks. We used the bag of word (BoW) representation with term frequency, and term frequency-inverse document frequency(TF-IDF) features to obtain the numerical feature matrix for the text. Bag of word model provides a feature matrix of the word counts in the vocabulary for all the job descriptions. While term frequency-inverse document frequency provides a weighted feature matrix for the vocabulary in the job descriptions where it gives higher weight for unique and more distinctive words and lower weight for the most occurring words. We used a probabilistic model which is Naive Bayes classifier, to build our two text classifiers, one with BoW and TF-IDF features. The models scored  87\% and 88\% for the models based on BoW and TF-IDF features, respectively.

\subsubsection {Advanced Methods}
Advanced Methods include models that were developed using deep learning-based techniques coupled with pre-trained word embeddings to build powerful text classifiers for our task. We represented the job descriptions using many pre-trained word embeddings to obtain our feature vectors for each job. We used state of the art pre-trained word embeddings that includes Elmo, BERT, RoBERTa, TransformerXL, and a average of Bert, RoBERTa, and ELMo embeddings that is called stacked embedding. Furthermore, to build our text classification models, we used the flair framework. Flair framework provides an implementation for almost all NLP tasks and encapsulates various state of the art pre-trained embedding\footnote{\url{https://github.com/flairNLP/flair}}. The models were trained using recurrent neural network (RNN) and used the Cross-Entropy Loss as the loss function. We also built an ensemble model based on the models developed. The ensemble model outputs the label that had the highest probability among the various models.  Table \ref{mod_accuracy} describes the results of the multiple models created.

\subsubsection {Validation}
\subsubsection{Manual labels}
Since DoT and O*NET structures are different, we wanted to check the feasibility of using O*NET based classifier on labeling DoT job descriptions. From each DoT edition, a sample of almost 1\% (200 jobs from each edition) of the data was extracted and manually labeled by human annotators.  The human annotators categorize the job according to their understanding of the job description's content. The table describes the results of the models when testing them with the manually labeled sample. The resulting accuracy is high indicating that our classifier is valid in classifying DoT descriptions.

\subsubsection{Metadata Validation}
\begin{figure}[h!]
    \centering
    \includegraphics[scale=0.8]{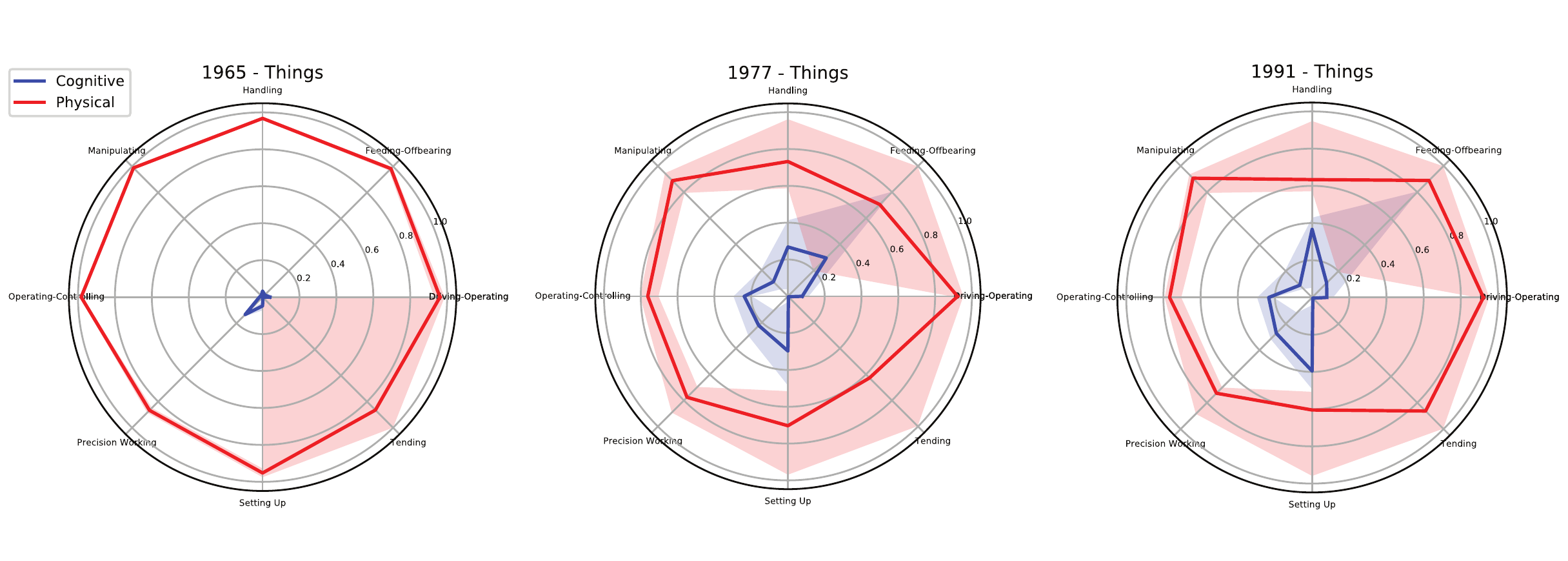}
    \caption{Radar chart showing the percentage of jobs classified as physical (red) and cognitive (blue) on Things for 1965 ,1977 and 1991. Some worker functions has a high confidence intervals that the others due to lack of data points for them.}
    \label{fig:Things}
    \end{figure}

Starting from 1965, DoT started including metadata information in occupational codes to indicate skill requirement. The fourth, fifth, and sixth digits of the occupational code reflect the job’s relationships to Data, People, and Things, respectively. Each category encapsulate work functions or skills related to the direct interaction of the worker and that skill. We used these categories to validate the results of our highest performing classifier. For each classifier label, we aggregated the number of jobs in each worker function then calculated the percentage of jobs that reflect each label to see if a worker function is leaning towards the physical or the cognitive cluster. We also checked the robustness of the results via bootstrapping a 10,000 samples. We found that the category of ‘Things’, in Figure \ref{fig:Things},provide a some noise in certain worker functions due to sparsity of the data in them.

\subsubsection{Classification Based on The Job Title }
After looking into the results of the classifiers, we noticed that some jobs were labeled incorrectly with a high probability. After investigating the descriptions, we found that the description of some jobs that are cognitive describe some physical tasks. For example, the description of SUPERVISOR, BRIDGES AND BUILDINGS is "Directs and coordinates activities of workers engaged in construction and repair of railroad structures, such as bridges, culverts, tunnels, and buildings: Plans work schedules for construction and maintenance projects.. etc." include a description of workers under the supervision. So we defined a list of frequent words in the job titles and modified the labels for all the jobs that include those keywords in their titles. The keywords include "OPERATOR" and "MAKER" that is directly classified as physical and "SUPERVISOR" and "MANAGER" that is directly classified as cognitive. We tested this modification with the human labels, and the accuracy of the classification increased by one percent.

%\begin{table}[ht]
%  \begin{tabularx}{\linewidth}{XX}
%    \toprule Model & Accuracy\\
%    \midrule
%    Naive Bayes - BoW                      & \hl{X}\%      \\
%    Naive Bayes - TF-IDF                     & \hl{X}\%      \\
%    ELMO                      & \hl{82}\%      \\
%    BERT                     & \hl{83}\%        \\
%    RoBERTa                         & \hl{83}\%       \\
%    TransformerXL Embeddings & \hl{83}\%   \\
%    Stacked Embeddings(Bert, RoBERTa and ELMo) & 87\%   \\
%    Ensemble model & \hl{85}\%   \\
%  \bottomrule
%  \end{tabularx}
%  \caption{Summary of the validation results of the models}
%  \label{ev_accuracy}

%\end{table}

\section{Extended Results and Discussions}

\subsection{Jobs Networks and Classifications Over Time}
We derive a job network for each edition of the Dictionary of Titles where nodes are jobs, and edges are defined based on text similarity between job descriptions. The derived networks are in figure \ref{fig:part1} B. The networks provide a unique insight into the network structure of the US labor marketplace over a 50 year period. ForceAtlas2 algorithm was used for network specialization. Using the classification results of the model with the highest accuracy on the validation process, which is the stacked model, we labeled and projected the classification result for each job in DoT. In figure  \ref{fig:part2} B, we can visually see that the two clusters are becoming more polarized, and their jobs are getting more grouped overtime with an increasing number of cognitive jobs.

\subsection{Labor Polarization Over Time}
\begin{figure}[h!]
    \centering
    \includegraphics[scale=0.4]{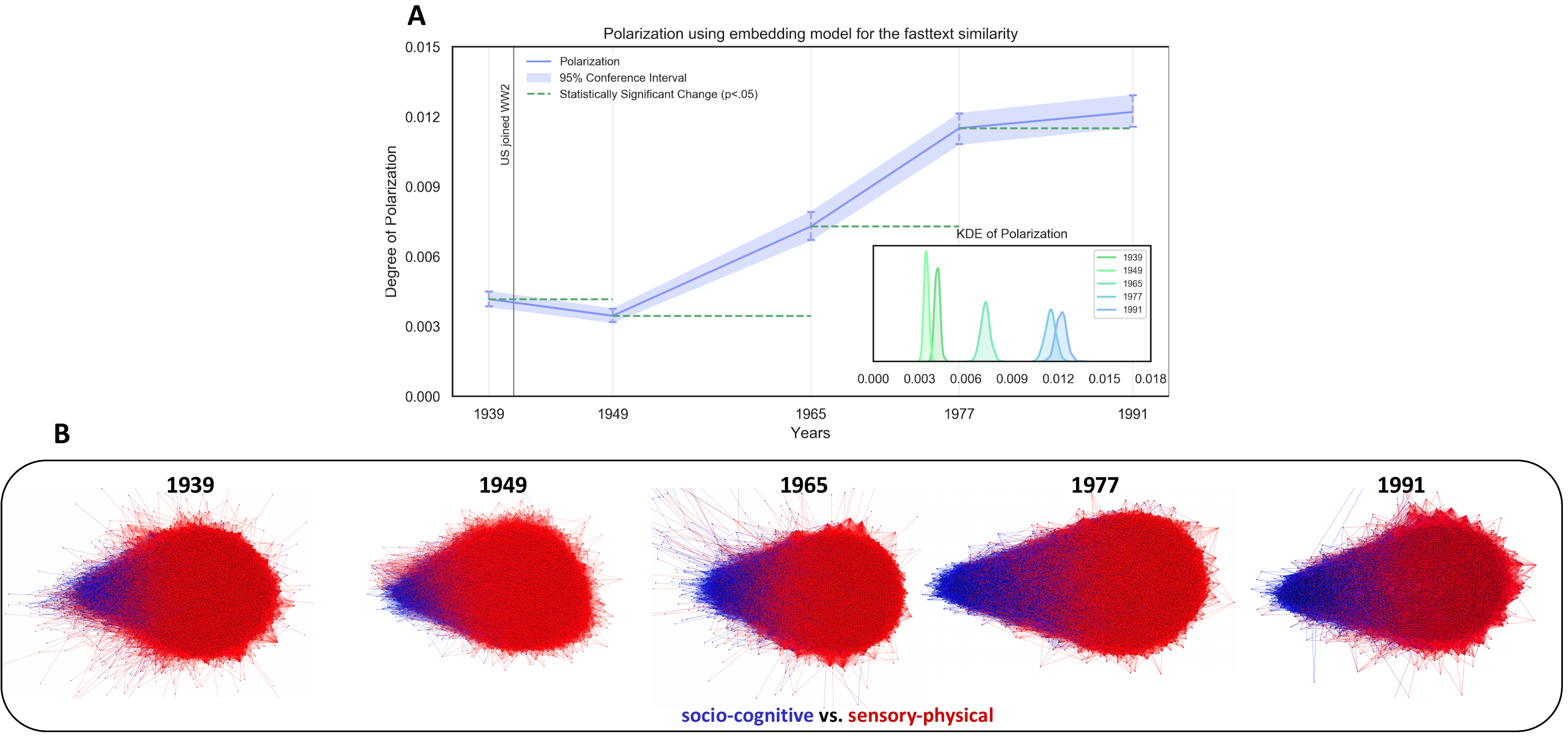}
    \caption{Network polarisation}
    \label{fig:part2}
\end{figure}

The modularity of the resulting networks gives a quantitative measure of job polarisation and changes overtime. As described in section \ref{polarization}, we want to measure and explore if the density of the connections within the cognitive cluster is increasing overtime as opposed to its connection with the physical cluster and vise versa. This increase in modularity indicates that job polarization has generally been increasing overtime. Our calculations confirm that, as seen in figure \ref{fig:part2}A. In addition, we tested the statistical significance of the results and found that there is a statistically significant difference between the polarization of each edition, as shown in the green lines in the figure. We tested the polarization using different cluster partitions based on the multiple classifiers we developed and obtained the same reported trend, the polarization figures are in the SI.

    \begin{figure}[h!]
    \centering
    \includegraphics[scale=0.6]{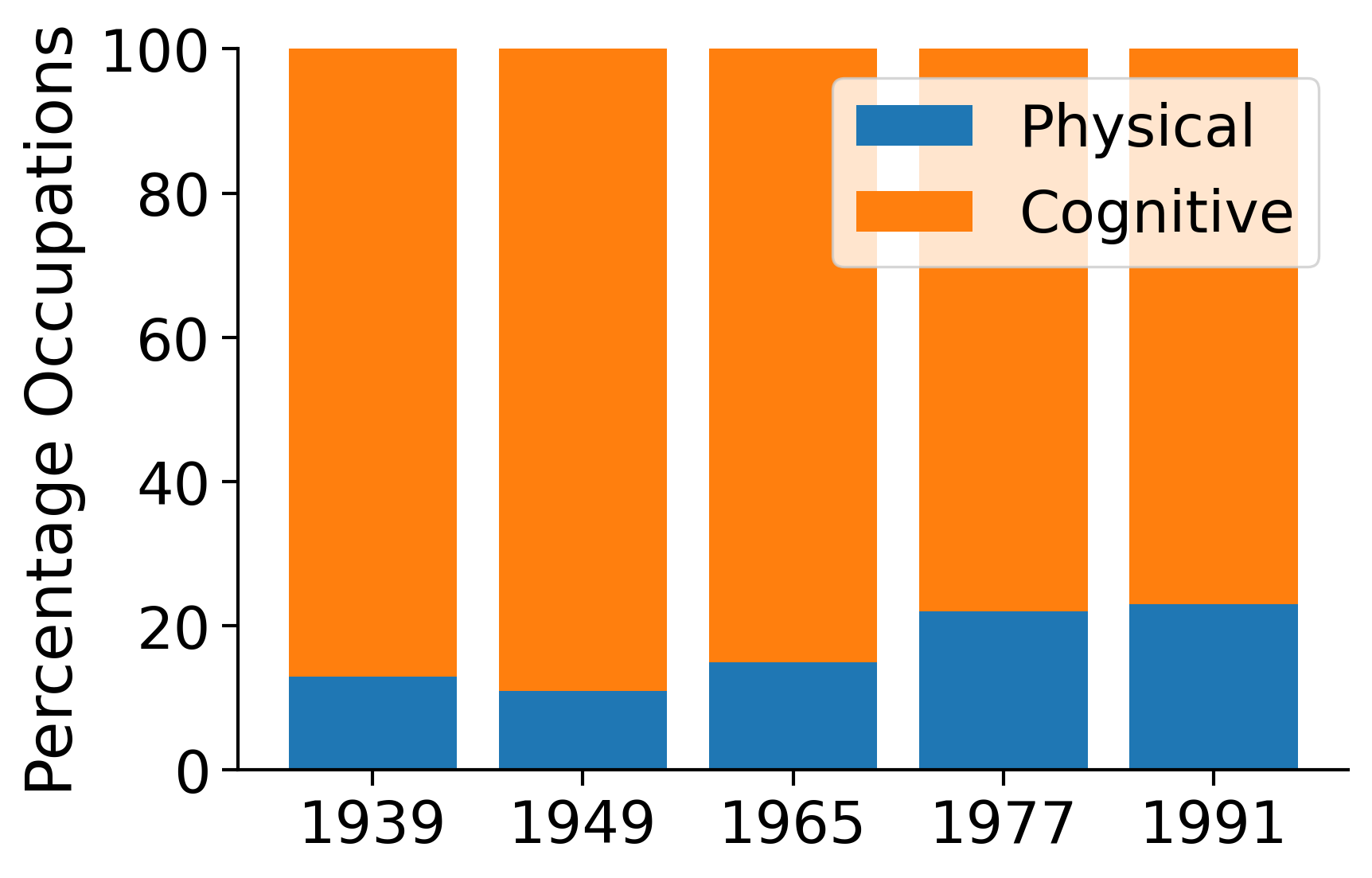}
    \caption{Proportions of occupations labelled as physical and cognitive in each edition of the DoT}
    \label{fig:phys_cog_percentage}
\end{figure}

\begin{figure}[h!]
    \centering
    \includegraphics[scale=0.6]{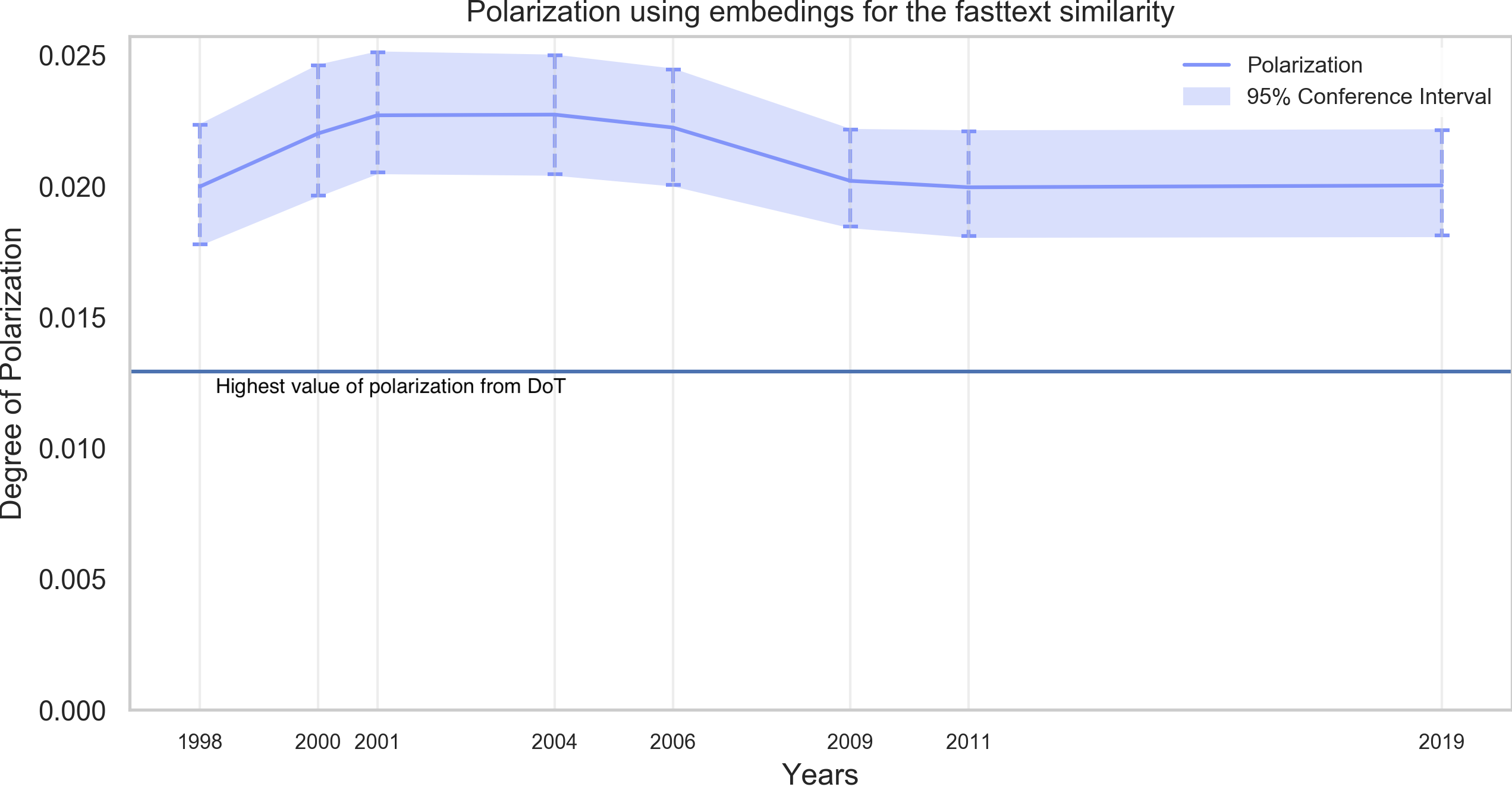}
    \caption{The Degree of Polarization in the 21 Century}
    \label{fig:onet_pol}
    \end{figure}

    \begin{figure}[h!]
    \centering
    \includegraphics[scale=0.6]{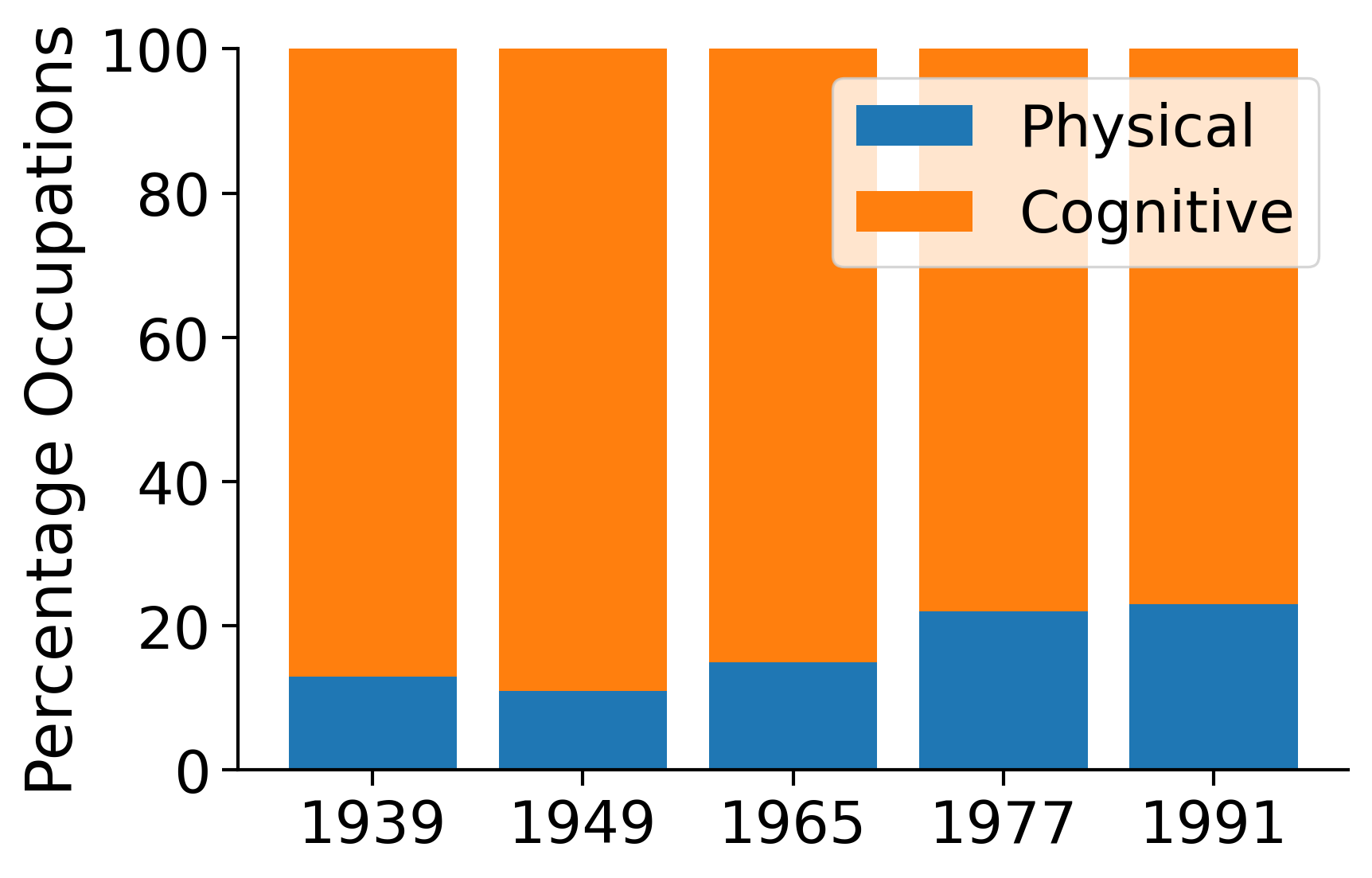}
    \caption{Proportions of occupations labelled as physical and cognitive in each edition of the DoT}
    \label{fig:phys_cog_percentage}
\end{figure}

\subsection{Polarization in the 21st Century }
We also applied the same methodology to find the degree of polarization in the twenty-first century using O*NET data that extends annually from 1998 and continued onward. We calculated the similarity of job descriptions in O*NET in the duration of from 1998 to 2019 and  notices that the descriptions in O*NET is not updated as frequent it is only updated seven times. Figure \ref{fig:onet_pol} shows that the polarization in the twenty-first century is higher than the one obtained from DoT. We know that during the period of the tech bubble (1995-2001) there had been an increase of employment in the technology sector and also an increase in college enrollments in computer science related fields. That could be the reason that there has been an increase in polarization in the first three releases of O*NET. We also assume that the great recession has affected the polarization leading it to a decline in the period 2007-2011 and then stabilizes for the following releases of the database.

\section{Network Modularity Under Unbalanced Labels}

We consider the definition of modularity

\begin{equation}
    Q = \frac{1}{2m} \sum_{ij} \Big[A_{ij} - \frac{k_{i} k_{j}}{2m}\Big] \delta_{c_{i},c_{j}}
    \label{moddef}
\end{equation}

Where $m=kn^{2}$ is the total number of edges, $A$ is the adjacency matrix, $k_{i}$ is the degree of node i and $\delta_{c_{i},c_{j}}$ is 1 for nodes of the same class and 0 otherwise.\\

The modularity of a network is thus determined by both (i) the labels or classes assigned to each node and (ii) the detailed edge structure, or more specifically, the extent to which edges are more likely to be found between nodes of the same class. Typically, the edge structure of a network is fixed and classes are assigned to nodes, as an iterative optimisation process, in order to best partition the network by maximising the modularity.\\

However, in the case of the occupation networks presented in the main paper, the edge structure \textit{and} the node classes (i.e. the job is either physical or cognitive) are both assigned. Thus we measure the emergent modularity and do not (and cannot) alter either (i) or (ii) in order to maximise the measured modularity.\\

We note that the degree of imbalance between classes (we consider only a binary classification appropriate for our use case, but this could be extended generally) affects the network modularity, with all other properties held constant. We demonstrate this by constructing a regular grid network with a fixed degree $k$. We assign nodes to one of two classes randomly, with the balance between the two classes being an independent variable ($p_{0}=(1-p_{1}$). We find that the modularity is minimised when the classes are balanced and maximised when one is dominant.

\begin{figure}[h!]
    \centering
    \includegraphics[scale=0.9]{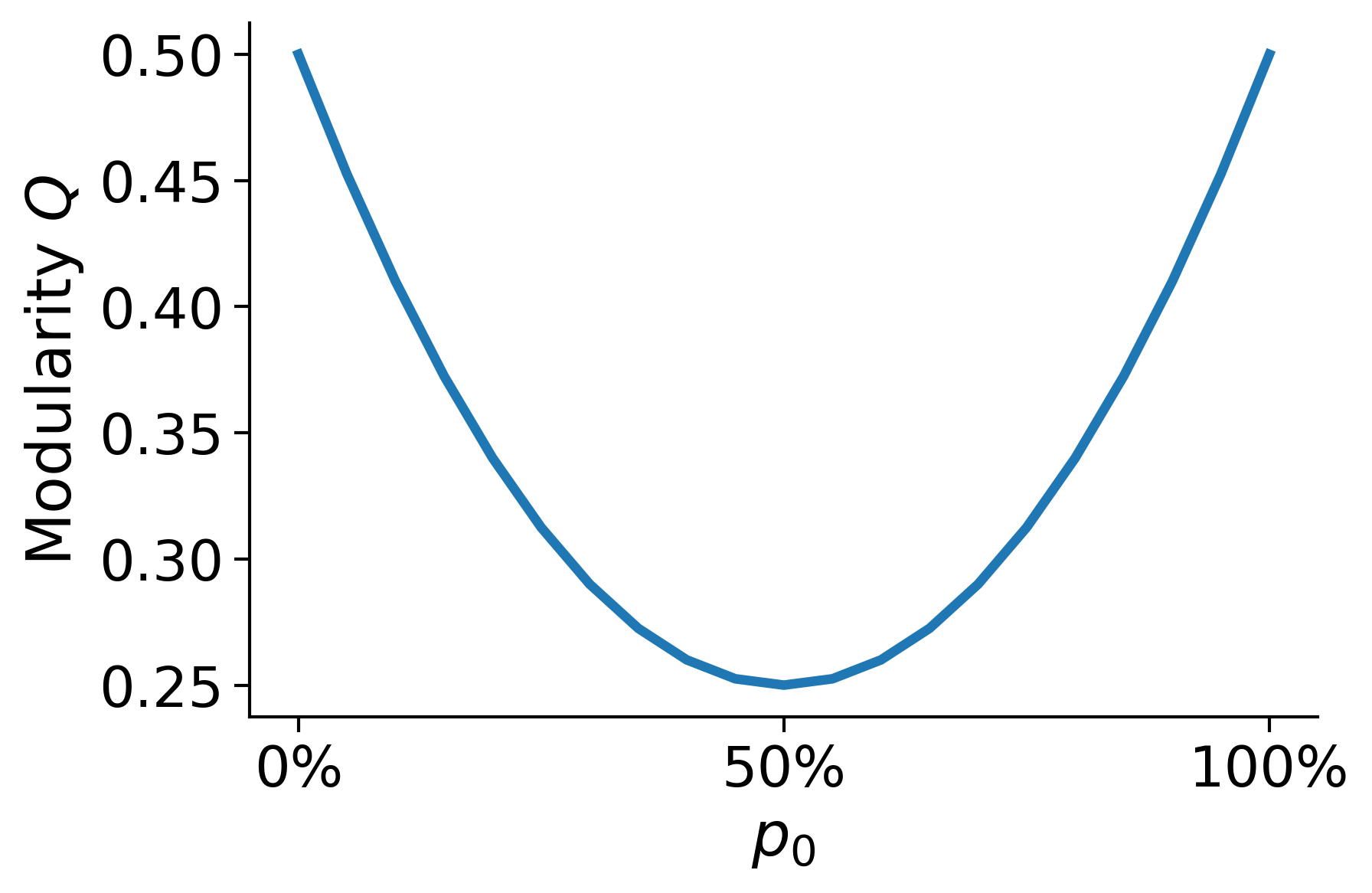}
    \caption{Modularity as a function of class imbalance in a regular grid network}
    \label{fig:proportion}
    \end{figure}

We can simplify (\ref{moddef}) in the special case of a regular grid. In this case the term in $k$ can be removed from the sum.

\begin{equation}
     Q = \frac{1}{2m}  \Big[A_{ij} - \frac{k_{i} k_{j}}{2m}\Big] \sum_{ij} \delta_{c_{i},c_{j}}
\end{equation}

The sum over the delta function can be expressed in terms of the class proportions $p_{0}=(1-p_{1})$. The number of node pairs of the same class is given by $n^{2}(p_{0}^{2} + p_{1}^{2})$. Substituting for $k$ we find

\begin{equation}
    Q = \frac{1}{2k} \Big( k-\frac{k}{2n^{2}}\Big) (p_{0}^{2} + p_{1}^{2})
\end{equation}

When the edge density is low ($k <<n^{2}$) this simplifies to

\begin{equation}
    Q = \frac{1}{2} (p_{0}^{2} + p_{1}^{2}) = \frac{1}{2} (p_{0}^{2} + (1 - p_{0})^{2})
\end{equation}

We note that this form is independent of the size of the network ($n$) and also the mean degree ($k$). When only one class is present ($p_{0}=1$) $Q = 0.5$.\\

When comparing the 5 editions of the DoT, we have a varying class imbalance. We would like to disentangle the contributions to the measured modularity that are due to (i) class imbalance and (ii) edge structure. Therefore we report the empirical modularity relative to the modularity in a regular grid network \textit{with the same class imbalance}. This effectively controls for changing proportions of physical and cognitive jobs between editions and isolates the contribution to modularity due to edge structure and a preferential connection between nodes of the same class.

\begin{equation}
    \bar{Q}(p_{0}) = \frac{Q(p_{0})}{Q^{rand}(p_{0})}
\end{equation}

%\printbibliography
%\bibliography{references.bib}